\documentclass[sigconf, nonacm]{acmart}

\usepackage{graphicx}
\usepackage{caption}
\usepackage{subcaption}
\usepackage{hyperref}
\usepackage[compact]{titlesec}
\usepackage{multirow}

\usepackage{algorithm2e}
\usepackage{algpseudocode}   
\usepackage{latexsym}
\usepackage{url}

\begin{document}

\title{Efficient Representation of Interaction Patterns with Hyperbolic Hierarchical Clustering for Classification of Users on Twitter}

\author{Tanvi Karandikar}
\authornotemark[1]
\affiliation{
  \institution{International Institute of Information Technology, Hyderabad}
%   \city{Hyderabad}
%   \state{Telangana}
  \country{}
}
\email{tanvi.karandikar@students.iiit.ac.in}

\author{Avinash Prabhu}
\authornote{Authors contributed equally to this research.}
\affiliation{
  \institution{International Institute of Information Technology, Hyderabad}
%   \city{Hyderabad}
%   \state{Telangana}
  \country{}
}
\email{avinash.prabhu@students.iiit.ac.in}

\author{Avinash Tulasi}
\affiliation{
  \institution{Indraprastha Institute of Information Technology, Delhi}
%   \city{Delhi}
  \country{}
}
\email{avinasht@iiitd.ac.in}

\author{Arun Balaji Buduru}

\affiliation{
  \institution{Indraprastha Institute of Information Technology, Delhi}
  \country{}
}
\email{arunb@iiitd.ac.in}

\author{Ponnurangam Kumaraguru}
\affiliation{
  \institution{International Institute of Information Technology, Hyderabad}
%   \city{Hyderabad}
%   \state{Telangana}
  \country{}
}
\email{pk.guru@iiit.ac.in}

\begin{abstract}

    Social media platforms play an important role in democratic processes. During the 2019 General Elections of India, political parties and politicians widely used Twitter to share their ideals, advocate their agenda and gain popularity. Twitter served as a ground for journalists, politicians and voters to interact. The organic nature of these interactions can be upended by malicious accounts on Twitter, which end up being suspended or deleted from the platform. Such accounts aim to modify the reach of content by inorganically interacting with particular handles. These interactions are a threat to the integrity of the platform, as such activity has the potential to affect entire results of democratic processes. In this work, we design a feature extraction framework which compactly captures potentially insidious interaction patterns. Our proposed features are designed to bring out communities amongst the users that work to boost the content of particular accounts. We use Hyperbolic Hierarchical Clustering (HypHC) which represents the features in the hyperbolic manifold to further separate such communities. HypHC gives the added benefit of representing these features in a lower dimensional space -- thus serving as a dimensionality reduction technique. We use these features to distinguish between different classes of users that emerged in the aftermath of the 2019 General Elections of India. Amongst the users active on Twitter during the elections, 2.8\% of the users participating were suspended and 1\% of the users were deleted from the platform. We demonstrate the effectiveness of our proposed features in differentiating between regular users (users who were neither suspended nor deleted), suspended users and deleted users. By leveraging HypHC in our pipeline, we obtain F1 scores of upto 93\%. 
    
    % We show that HypHC performs better than other established dimensionality reduction techniques in terms of classification performance. Experiments with multiple classifiers on different class separation configurations show the effectiveness of our features and corroborate the benefits of HypHC as a dimensionality reduction strategy.
  
\end{abstract}

\maketitle

% TODO: make sure you include. 
% \begin{IEEEkeywords}
% Knowledge Graph; Uncertainly requirement Analysis; Multi-round dialogue; Cognitive Service Computing; chat-bots; Conversational AI Bot; Granular Computing.
% \end{IEEEkeywords}

% \fi

% \IEEEpeerreviewmaketitle

    % \cite{le2019postmortem, chowdhury2021examining, chowdhury2020twitter} 
    
\vspace{-1.5ex}
\section{Introduction}\label{sec:intro}
    Discussions on Online Social Networks (OSNs) are a key player in huge democratic processes such as elections. Recently, the 2020 U.S. General Elections \cite{chowdhury2021examining} 
    and the subsequent Capitol Riots \cite{capitol_news} have been heavily influenced by conversations on OSNs such as Twitter and Parler \cite{prabhu2021capitol}. Politicians and political handles are very active on OSNs, especially during times of democratic elections
    \cite{elections_increase}. The nature of their engagement on these platforms is often an important part of their campaign strategies
    \cite{barack}. Particularly, OSNs have played an important part in the 2014 and 2019 General Elections in India, where studies show the success of the winning party was closely associated with their use of Twitter to engage with voters \cite{Ahmed2016, saurabhh}. Users of these OSNs can interact with the content shared by political parties. The more interaction such content gets, the higher the popularity and higher the chance that such content reaches more people
    \cite{Boulianne2021}. Higher engagement with posts can also lead to benefits such as a higher number of followers and more media coverage \cite{Keller2018}. This engagement could lead to a domino effect that can turn the tide of election results \cite{mobilizing_trump}. Due to the effect of this engagement on the outcome of democratic processes, it is important for OSNs to be fair in the way content is boosted.

    Twitter in particular aims to keep their algorithm fair by endorsing and boosting content that is of genuine interest to most users.\footnote{https://help.twitter.com/en/rules-and-policies/platform-manipulation} This is important to ensure its status as a reliable platform for all political parties as well as voters by preserving interests spanning across all users.\footnote{https://help.twitter.com/en/rules-and-policies/election-integrity-policy} However, a group of users can manipulate the system by closely working together to artificially engage with the posts of a particular account.\footnote{https://www.cigionline.org/articles/how-bjp-used-technology-secure-modis-second-win/} During democratic processes like elections, collusive behaviour by users can be a serious threat to the integrity of a platform  \cite{le2019postmortem}. Identifying and moderating such malicious accounts thus becomes critical for these social media giants. 
    
     Accounts which are found to be participating in malicious engagement with posts with an intent to artificially boost their reach can be suspended from Twitter. The presence of such accounts could constitute large proportions of engagement with a politician's account \cite{Onuchowska2019RocketSO}. 
    Along with suspended accounts, accounts that are ultimately deleted from the platform can also pose a threat. Media has reported the spread of misinformative and deceptive content by fraudulent accounts that end up being deleted from the platform \cite{runet}. Accounts that end up being suspended or deleted can form significant portions of the aforementioned collusive groups \cite{Onuchowska2019RocketSO}, and thus identifying such accounts becomes very important, especially in the context of democratic processes like elections. 
        
    During the Indian General Elections of 2019, we observed three major classes of users. 1) Regular users, 2) Suspended users and 3) Deleted users. The class \textit{suspended} consists of those users who were identified to be violating the Twitter policy and were suspended by the Twitter platform. The class \textit{deleted} consists of users who were once active on Twitter but whose accounts were later deleted. \textit{Regular} user accounts are Twitter users who were neither suspended nor deleted from the platform.

    In the real world, any given OSN itself does not have information on the motivation behind a user's behaviour. The only information such platforms have regarding the users is the engagement of the users with the platform. To distinguish between the different classes of users, we make use of this same information available publicly. Given that Twitter is only able to observe user activity but not the ultimate outcome (here, suspension or deletion of the account), we design a set of features which are class independent. We then use these features to train models to distinguish between different classes of users.

      User interactions in OSNs often emulate trees with users forming communities \cite{ganea2018hyperbolic, ganea2018hyperbolicentailment, chami2019hyperbolic}. The communities of users which participate in artificial engagement can have deep hierarchies.$^3$ Capturing such extreme hierarchies in the traditional Euclidean space is hard and inefficient  \cite{ganea2018hyperbolic}. So, for our work we use hyperbolic manifolds \cite{ratcliffe1994foundations} which are geometric structures with a negative curvature. This representation space can also be perceived as a continuous version of a tree with area increasing as we move away from the origin (point of observation) of the plane. This nature of geometry is useful to capture the tree-like user communities \cite{hyphc_main}. 
     
       We use Hyperbolic Hierarchical Clustering - \textit{HypHC} as the hyperbolic representation technique to capture communities and reduce dimensionality \cite{hyphc_main}. HypHC provides the inherent advantage of community detection with information encoding. The reduced dimensions in the hyperbolic manifold ensures a computationally efficient downstream task. Additionally, representations obtained using HypHC can be treated just like Euclidean embeddings. We reduce the dimensionality of our features by a factor of ten without compromising on performance by leveraging hyperbolic manifolds via HypHC, all the while keeping every other component in our classification pipeline unchanged from its original Euclidean implementation.  

    The contributions of our work are as follows:
        \begin{enumerate}
            \item A feature engineering framework that can help OSNs engineer efficient representations which capture the interaction patterns with top handles. 
            \item Demonstrating the application of hyperbolic manifolds on real world social media data to reduce the computational and space complexity while effectively separating the regular, suspended and deleted accounts. 
        \end{enumerate}

\section{Related Work}
{

Our work covers two major domains: the study around classification of accounts on Twitter; and hyperbolic manifolds and their applications.

    \subsection{Classification of Users in OSNs}
        Earlier works that studied spam on Twitter  \cite{mccord2011spam}  leveraged user characteristics such as number of followers, and tweet content to generate features. These features were used to train a random forest classifier to detect spamming accounts. Follow-up works \cite{lee2014early} aimed to identify malicious accounts created in a short period of time by using account names. They compare algorithm-made names with man-made names by clustering accounts sharing similar name-based features. The work by Wei et al. \cite{wei2016exploring} uses temporal sentiment analysis to differentiate suspended users from non-suspended users with the help of statistical techniques like Naive Bayes classifier and SVM. 
        % They used generalised lexicons that combined several lexicons found in JST and VADER models to reflect the sentiment of users. 
    
        With claims that Twitter had been influencing voter sentiment during the U.S. elections, there have been focused attempts at characterizing Twitter's part in democratic processes in many nations 
        % \cite{doi:10.1177/1461444812470612, knight2012journalism, dzisah2018social, fadillah2019social}
        \cite{doi:10.1177/1461444812470612, knight2012journalism, dzisah2018social}. 
        % Authors in \cite{lee2019social} study the political influence of social media on the general public, and they conclude that the effect, if any, is negative. 
        Notable studies on characterizing users based on Twitter's moderation decisions \cite{le2019postmortem} show that the malicious communities suspended by Twitter exhibit a considerable difference from regular accounts. This class of works combines Twitter and elections to understand different user groups on the platform, spread of sentiment and news on the network and echo chambers and their effects, to minimize the effect Twitter has on democratic processes. In other works \cite{chowdhury2021examining}, authors group Twitter users into communities based on their retweet and mention networks and analyze different characteristics such as popular tweeters, domains, and hashtags. They found that malicious and regular accounts participate in communities which exhibit significant differences in terms of popular account and hashtag usage. 
        
        Work has also been done on identifying and characterizing deleted users on social media platforms. Authors in \cite{Bastos2021} found that a significant proportion of accounts involved in political discourse on Twitter revolving around the \textit{Brexit} referendum campaign were deleted from the platform. Volkova et al. used profile, network and behavior clues, sentiment and emotion features, text embeddings and topics to detect accounts deleted from Twitter which were active in the context of the Russian-Ukranian crisis \cite{runet}. Twitter itself removed over 2 million accounts from the platform which were suspected to be fake. These accounts were allegedly giving misleading follower counts for some accounts \cite{twitter_fake}.      
        
        % They used SVMs with social, lexical and user features to build a classifier. They also discuss the deletion of accounts used to spam Twitter. Following up in the same context, the work \cite{zhou2016tweet} establishes the "proper" use of Twitter in long sustained users. They use richer user features to include findings from the work \cite{petrovic2013wish}.
        
        % While majority of the works concentrate on characterizing different aspects of users behaviour on the platform, we aim to systematically study the policy that dictates \textit{welcoming / harmonious} user behaviour, and use the same to design features which reflect malicious intentions.

        % A characterization study \cite{bhattacharya2016characterizing, volkova2017identifying} use psycho linguistic features to find that regrettable Tweets with certain words get deleted with higher probability. However, democratic processes leads to agendas and the motivations of users to engage on OSNs change. For example, swearing or posting embarrassing tweets will have minor effects on user accounts compared to fake news propagation. While our work builds a classifier, we have a completely different setting of the Indian General Elections 2019. Our work differs from the past works significantly in context, majority of the above mentioned research is done on users in a casual environment with no or a wide context. We also aim at making computations efficient by lowering the dimensions of representations. In the next subsection we discuss Hyperbolic manifolds which are a key to faster decision making. 

    \subsection{Hyperbolic manifolds and applications}
    {
        Hyperbolic Hierarchical Clustering (HypHC) \cite{hyphc_main} is a key component of our work. We use this technique to reduce the dimensionality of our user representations for efficient computation. The following sub-section introduces previous research work on hyperbolic manifolds and their applications. 
        
        In the work \cite{nickel2017poincar} authors claim that the ability of embeddings to model complex patterns is bounded by the dimensionality of the embedding space, because of which it is impossible to extract embeddings of large graph-structured data without any loss of information. So, to increase the representation capacity of embedding methods, they used hyperbolic spaces. Representations in hyperbolic spaces are capable of effectively capturing the underlying hierarchical structures in data at lower dimensions \cite{feng2020hme}.
        
        Feng et al. \cite{feng2020hme} leveraged this property of hyperbolic spaces to build a hyperbolic metric emebedding model, which projects location-based check-in data of users from social media into the hyperbolic space to predict the next POI (point of interest) for a user. 
        Wang et al. \cite{9322242} proposed a hyperbolic geometry representation learning model to link user identities across different social network platforms. 
        
        Hyperbolic manifolds have also been used to embed knowledge graphs \cite{wang2020h2kgat}, images \cite{khrulkov2020hyperbolic} and words \cite{tifrea2018poincar} in far lower dimensions. Question-answering \cite{tay2018hyperbolic} and clustering \cite{monath2019gradient} models have also leveraged hyperbolic manifolds to improve their performance.  

    }
}

% \TK{Making a table for the timeline}

\section{Data description}
{
    In this section we introduce and describe the dataset used and the definitions of the different classes of users in our dataset, i.e. regular, suspended and deleted users. 

    For our work we use the `Analysis of General Elections 2019 in India' (AGE2019) dataset \cite{AGE2019}. This dataset consists of tweets that span from February 5th 2019 to June 25th 2019. This covers a time period starting from two months prior to the first polling in the elections upto one month after the results of the elections were declared. The tweets are collected by querying the Twitter data collection APIs to retrieve tweets having hashtags related to the 2019 Indian General Elections. A total of 45.6 million tweets made by 2.2 million unique users are collected in the dataset.
    
    The AGE2019 dataset also consists of two lists of user ids - deleted users and suspended users. These are users who were found to be either deleted or suspended from the platform as of June 29th, 2019. A total of 56,927 of these users were identified to be suspended as they returned error code \textit{63} on querying the Twitter API. These users form our \textit{suspended} class. Additionally, 21,083 users returned error code \textit{50}, signifying that these accounts had been deleted from the platform. These users form our \textit{deleted} class. The AGE2019 dataset also shared a list of 100,000 users randomly sampled from the remaining users (neither suspended nor deleted). This set of users forms the \textit{regular} class of users in our study. To generate features to separate the deleted, suspended and regular classes of users, we use the election-related tweets by each user, that are provided as a part of the AGE2019 dataset.
}

\section{Feature Engineering}\label{sec:featureenginering}
{
   \begin{figure}
        \centering
        \includegraphics[width=\linewidth]{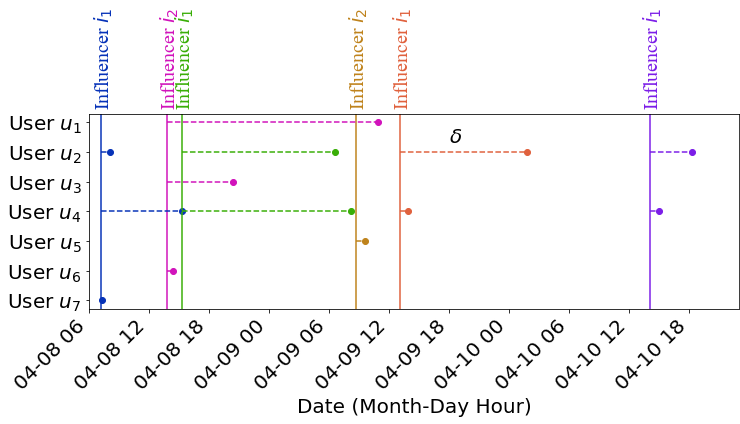}
        \caption{Interaction patterns between the top two Influencers and seven randomly sampled users over a two-day time period. X-axis (bottom) shows the time stamps of the tweets and retweets, Y-axis (left) shows the users. X-axis (top) shows the Influencers. Each vertical line represents a tweet by an Influencer. Each dot in the graph indicates a retweet of an Influencer's tweet by a user, joined by a dashed line to the original tweet. The length of each dashed line represents the time delay $\delta$ between the Influencer's original tweet and the user's retweet of that tweet.}
        \label{fig:collusion_fig}
    \end{figure}

    In this section, we describe our feature engineering process. Past research works have used features derived from the content of the tweets, extensive graphs of interactions between the users and more to extract features like sentiment, emotion, lexical features etc. to distinguish between classes on Twitter \cite{volkova2017identifying, runet}. However, these can become complicated to extract given the large amounts of data involved and complicated multi-lingual nature of the tweets. We use features that can easily be extracted based on information captured from the user's profile and tweet activity, and do not delve into the actual content of the tweets. Section \ref{sec:comparison_user_level} demonstrates the importance of our features by comparing the same with some \textit{user-level features} that past works have used to model the account characteristics. We use this section to explain the motivation and design of our proposed \textit{interaction features} to capture the nature of interactions of each user with top profiles that emerged during the General Elections.

    \subsection{Interaction features}
    {
    \label{sec:interaction_features}
    
        During the 2019 General Elections in India, contesting parties have been known to maintain an \textit{IT cell} \cite{logicalindian, campbell2019global}. These IT cells, along with a dedicated team of supporters work to push the agenda of their parties as much as possible. As mentioned in \cite{campbell2019global}, these IT cells are known to propagate particular agendas on OSNs by engaging with posts by a certain account. So, artificial engagement in the context of the General Elections revolves around boosting the popularity and pushing the ideologies of a particular leader or party. 
        
        To identify the top leaders, we curate a list of users whose content is widely shared across the platform. We call them \textit{Influencers} in our work because of the effect these users have on the content shared on the platorm. We particularly look at the engagement on the platform as a result of their tweets. Influencers need not be political leaders, but in the context of the 2019 General Elections, we observed that most Influencers are political leaders or political party handles. We use these Influencers to generate our interaction features. 
        Table \ref{tab:notations} summarizes the notations used henceforth. 

\begin{table}[H]
        \centering
          \begin{tabular}{c|l}
          \hline
                \textbf{Notation} & \textbf{Meaning} \\ \hline
                
                $I = \{i_1, i_2, \dots \}$ & Set of all Influencers \\
                $p$ & Number of Influencers in $I$ \\
                $rt\_score$ & Number of users retweeting a particular \\ & Influencer\\
                $T_y = \{t_{y1}, t_{y2}, \dots\}$ & Set of all Tweets by Influencer $i_y$ \\
                $U = \{u_1, u_2, \dots\}$ & Users engaging with the Influencers in $I$\\
                $R_{xy}$ & All retweets by user $u_x$ of Influencer $i_y$\\
                $t_{yk}$ & $k$th Tweet by Influencer $i_y$ \\
                $r_{xt_{yk}}$ & User $u_{x}$'s retweet of tweet $t_{yk}$  \\
                $\delta_{xt_{yk}}$ & Delay in retweeting $t_{yk}$ by user $u_{x}$ \\
                $D_{xy}$ & Set of all delays $\delta_{xt_{yk}}$\\
                $\Delta_{xy}$ & Median of all delays in $D_{xy}$\\
                $n_{xy}$ & Number of times $u_x$ retweeted $i_y$ \\
                $\vec{V}_{x}$ & Interaction features for $u_x$ \\               \hline
                
                \end{tabular}
            \caption{Notations used in Section \ref{sec:interaction_features}, Algorithm \ref{algo} and Figures \ref{fig:plaw}, \ref{fig:framework}.}.
            \label{tab:notations}
        \end{table}      
        
        \newpage
        Retweets are a great way to quickly engage and amplify the reach of a tweet. Those retweets without any text of their own are just endorsements \cite{rt_are_endors, cork2017retweet, kim2012role}. We take all retweets from the data and curate a list of user profiles $I'$ whose tweets have been retweeted. For each such user in $I'$, we keep track of how many users in the dataset retweeted their tweets (i.e. $rt\_score$). We then consider the top $p$ user profiles having the highest number of retweets as Influencers to form the set $I$.

        The set of users interacting with Influencers in $I$ is given by $U =  \{u_1, u_2, \dots\} $ where each user in $U$ has engaged with at least one Influencer in $I$ by retweeting at least one of their tweets. For each Influencer, we define $T_y = \{t_{y1}, t_{y2} \dots\}$ to be the set of all tweets by Influencer $i_y$. For each pair $(u_x, i_y)$ where user $u_x \in U$ and Influencer $i_y \in I$, we look at those tweets by user $u_x$ which are retweets of any tweet by Influencer $i_y$, i.e. retweet of any tweet in $T_y$. We thus have a set of retweets
        $R_{xy} = \{$set of all $r_{xt_{yk}}\} $ where $r_{xt_{yk}} $ is a retweet of Influencer $i_y$'s tweet $t_{yk} \in T_y$ by user $u_x$. If a user $u_x$ has never retweeted a tweet by Influencer $i_y$, then $R_{xy} = \{\}$. In order to quantify a particular user's ($u_x$) interaction with any Influencer ($i_y$), we use two values:

        \begin{enumerate}
        \item The first value we use is the delay, or time lag in retweeting.
        For each retweet $r_{xt_{yk}} $ in $R_{xy}$, we define the corresponding delay $\delta _{xt_{yk}}$ to be the absolute value of difference in seconds between time of the original tweet ($t_{yk}$) and the time of retweet of that tweet ($r_{xt_{yk}} $), i.e. $\delta _{xt_{yk}} =$ (time of retweet $r_{xt_{yk}}$ -  time of original tweet $t_{yk}$). Figure \ref{fig:collusion_fig} shows an illustrative demonstration of how each such $\delta$ is calculated. Thus, for each pair $(u_x, i_y)$ we have a set $D_{xy}$ which is the set of all $\delta_{xt_{yk}}$. $D_{xy}$ forms the set of delays for each retweet by user $u_x$ of Influencer $i_y$. We take the median of the delays in $D_{xy}$ to get the final delay $\Delta_{xy}$. 
        
        \item The second value that we use is the number of times the user has retweeted that Influencer's tweets. We define $n_{xy}$ to be the number of elements in $R_{xy}$, i.e. the number of times user $u_x$ retweeted a tweet by Influencer $i_y$. 
        \end{enumerate}
        Thus for each pair $(u_x, i_y)$, we have a two-element vector $\vec{v}_{xy}$ which has two values: delay $\Delta_{xy}$ and the number of retweets $n_{xy}$. We form the interaction feature vector $\vec{V}_x$ for each user $u_x$ by concatenating all such vectors $\vec{v}_{xy}$ obtained for each corresponding Influencer $i_y$ where $i_y \in I$. In case a user has never interacted with an Influencer (i.e. $R_{xy} = \{\}$), we set $\Delta_{xy}$ to a large negative value, and $n_{xy}$ to 0. We choose a large negative $\Delta_{xy}$ in this case because such a value would never appear if $R_{xy}$ was not empty, thus achieving a good separation in the feature space. Algorithm \ref{algo} describes the above explained feature engineering process in a pseudo-code format. 
              \begin{algorithm}[h!]
        \KwData{AGE2019: $Tweets$ (all tweets) and $Users$ (all users)}
         \KwResult{Interaction features of all users }
         \hrulefill\\
         potential Influencers $I' \leftarrow \{\}$\;
         \ForEach{tweet $t$ in Tweets}{
            \If{tweet $t$ is a retweet}{
                \If{original poster $op$ of $t \notin I'$ }{
                    $I' \leftarrow I' + op$;
                }
            }
         }
         
         \ForEach{potential Influencer $i$ in $I'$}{
             $i$'s $rt\_score \leftarrow 0$; \\
             \ForEach{user $u$ in Users}{
                \If{user $u$ has retweeted any tweet by $i$}{
                    $i$'s $rt\_score \leftarrow i$'s $rt\_score + 1$\;
                }
             }
         }
         
         \hrulefill\\

         sort Influencers $I'$ by their $rt\_score$\;
         take top $p$ Influencers with highest $rt\_score$ as $I$\;
         $U \leftarrow \{\}$\;
         \ForEach{user $u$ in Users}{
            \If{user $u$ has retweeted the tweet of anyone in $I$}{
                $U \leftarrow U + u$\;
            }
         }
         \hrulefill\\
         
         \ForEach{user $u_x$ in $U$}{
            Feature vector $\vec{V}_x \leftarrow \{\}$;\\\
            \ForEach{Influencer $i_y$ in Influencers $I$}{
            $R_{xy} \leftarrow \{\}$; $D_{xy} \leftarrow \{\}$; \\
            let $T_y$ be the set of tweets by Influencer $i_y$\;
            \ForEach{tweet $t_{yk}$ in $T_y$}{
            \If{user $u_x$ has retweeted tweet $t_{yk}$}{
               {
                let $r_{xt_{yk}}$ be user $u_x$'s retweet of $t_{yk}$\; 
                % let $\delta _{{xt}_{yk}}$ be the delay for tweet $t_{yk}$\;
                $\delta _{{xt}_{yk}}$ =  time of $r_{xt_{yk}}$ - time of $t_{yk}$\;
                }
                 $R_{xy} \leftarrow R_{xy} + r_{xt_{yk}}$\;
                 $D_{xy} \leftarrow D_{xy} + \delta_{xt_{yk}}$\;
            }
            }
            $\Delta_{xy}$ = median of elements in $D_{xy}$\;
            $n_{xy}$ = number of elements in $R_{xy}$\;
            $\vec{v}_{xy} \leftarrow \{\Delta_{xy}, n_{xy}\}$;\\
            $\vec{V}_{x} \leftarrow \vec{V}_{x}$ + $\vec{v}_{xy}$
            }
         }
      \hrulefill
      \\
        \caption{Interaction features engineering. We first identify the Influencers $I$, then obtain the set of users $U$ who have interacted with them. We calculate the median delays and number of retweets of each user-Influencer pair to form the final feature vector.
        }
        \label{algo}
    \end{algorithm}  
       With the interaction feature vector $\vec{V}_{x}$, we quantify each user's interaction with the top handles. 
    %   If any account ended up being suspended or deleted due to reasons of targeted interaction with an Influencer account to inorganically increase its reach, our features will capture the same. 
      If there is a group of users colluding to interact with an Influencer account, their interaction feature vectors would look similar. 
      Thus our features will help to capture such groups of collusive accounts.
           }
   
    \begin{figure*}[ht!]
   \begin{subfigure}{0.32\textwidth}
     \includegraphics[width=\linewidth]{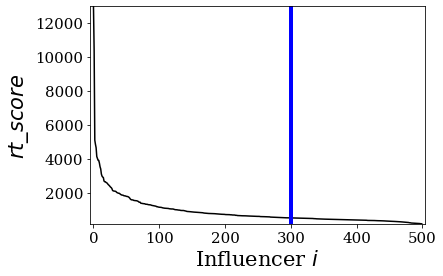}
     \caption{}
     \label{Fig:exp_a}
   \end{subfigure}\hfill
   \begin{subfigure}{0.32\textwidth}
     \centering
     \includegraphics[width=\linewidth]{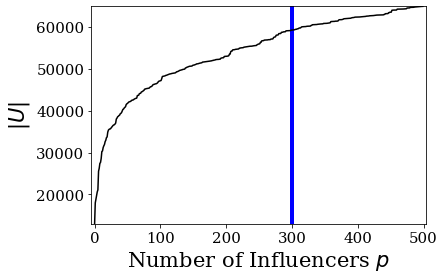}
     \caption{}
     \label{Fig:exp_b}
   \end{subfigure}\hfill
   \begin{subfigure}{0.32\textwidth}
     \raggedright
     \includegraphics[width=\linewidth]{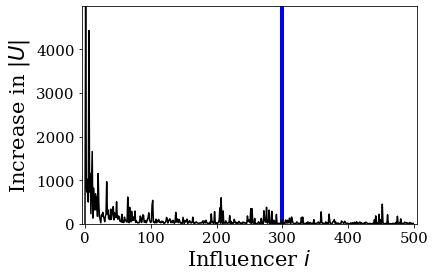}
     \caption{}
     \label{Fig:exp_c}
   \end{subfigure}
   \caption{Retweeting trends among users for the top 500 Influencers sorted in decreasing order of their $rt\_score$. Figure (a) depicts the number of unique users retweeting a particular Influencer $i$, i.e. $rt\_score$. Figure (b) is the number of unique users retweeting any of the top $p$ Influencers, i.e. the size of $U$ for a given size $p$ of $I$. Figure (c) shows the increase in $|U|$ for each new Influencer $i$ included in $I$. The vertical line in each figure represents our final choice of $p$ which is 300.}
    \label{fig:plaw}
\end{figure*}

     In order to generate interaction features we have to choose the number of Influencers $p$ to include in the set $I$. We observed the trends of number of unique users added to the set $U$ for each new Influencer added to the set $I$. We wish to choose $p$ high enough that a large number of users are incorporated. A larger set of users means different subgroups of these users interact with the tweets of different Influencers, thus capturing a larger number of collusive groups. However, at the same time we do not wish to have a large $p$ because we only want the top tweeting Influencers, i.e. the most popular ones, who have a relatively large number of users retweeting their tweets. In order to achieve this balance, we study the relationship between the number of Influencers and number of users as depicted in Figure \ref{fig:plaw}. In Figure \ref{Fig:exp_a}, each point on the x-axis represents a particular Influencer, and the y-axis depicts the number of unique users retweeting that particular Influencer (i.e. $rt\_score$). This indicates how many unique users are interacting with each Influencer. Higher the value, more popular the Influencer. 
    
    To get a sense of the diversity of users incorporated by choosing a particular $p$, we plot Figure \ref{Fig:exp_b}. In Figure \ref{Fig:exp_b}, the x-axis represents the number of Influencers $p$, and the y-axis represents the size of $U$ (or $|U|$) for a particular value of $p$. It is important to note that Figure \ref{Fig:exp_b} is not merely a cumulative plot of Figure \ref{Fig:exp_a} because there will be users who retweet the tweets of multiple Influencers. This graph is increasing, because as we add more Influencers to the set $I$, we incorporate more users in the set $U$. The amount of increase in the y-value of the graph as the x-value changes from ${p}$ to ${p+1}$ is the number of users added in $U$ when we increase the number of Influencers by one. To better visualise the effect of increase in $p$ on $|U|$ we plot Figure \ref{Fig:exp_c}.
    
    % A sharper slope upwards in Fig \ref{fig:plaw}.b at a particular point on the x-axis indicates a higher number of unique users incorporated when including that Influencer.  
    
    % To better visualise the effect of high slope in Fig \ref{fig:plaw}.b, we plot the difference in cumulative users retweeting Influencers \textit{1 to p} and \textit{1 to p-1}. 
    Each value in Figure \ref{fig:plaw}.c indicates the \textit{additional number of users} that are included for each new Influencer added to I. We reiterate that there will be users who retweet the tweets of multiple Influencers, and thus Figure \ref{Fig:exp_c} is different from Figure \ref{Fig:exp_a}. By increasing the value of our final chosen $p$, we include more spikes from Figure \ref{Fig:exp_c}, which means we incorporate a more diverse set of users in $U$. In Figure \ref{Fig:exp_c}, we observe that after around 300 Influencers, the \textit{additional number of users} that are included for each new Influencer added reduces. We hypothesize this to be the optimal value of $p$. To verify the validity of this hypothesis, we chose values of $p$ at intervals of 50 between 100 and 600 and found best results on our classifiers (the same classifiers described in Section \ref{sec:classifier}) with $p$=300. This confirmed our hypothesis. By choosing $p=300$ (as depicted with a vertical line in Figure \ref{fig:plaw}), we end up with $U$ containing 59,260 users. The class distribution of our final $U$ is depicted in Table \ref{tab:dist_table}. Note that we refer to our proposed user interaction features as $F$ henceforth.

        \begin{table}[H]
    \begin{tabular}{l|r}
    \hline
    \textbf{Class} & \multicolumn{1}{l}{\textbf{Number of Users}} \\ \hline
    Deleted & 8,078 \\
    Regular & 32,386 \\
    Suspended & 18,796 \\ \hline
    Total &  59,260
    \end{tabular}
    \caption{Class distribution of users after taking $p$=300 Influencers.}
     \label{tab:dist_table}
    \end{table}
    
    % \subsection{User-Level Features}
    % \label{sec:user_features}
    
}

\section{Dimensionality reduction using Hyperbolic Hierarchical Clustering}
{
        \label{hypclussec}
            Hyperbolic Hierarchical Clustering (HypHC) is a similarity based clustering method \cite{hyphc_main}. First, a binary tree with $n$ leaves is constructed. Each leaf node $i$ denotes a Twitter user that needs encoding to a lower dimension. From these leaf nodes, intermediate nodes that connect closer nodes are formed. If two leaf nodes are found to potentially belong to the same cluster, they have a least common ancestor (LCA). Each sub-tree denotes a potential cluster. The goal of HypHC is to cluster nodes in such a way that the pairwise similarity of the data is captured and used to form clusters. The binary tree is built such that the pair-wise similarity $sim_{i, j}$ between each pair of nodes $i, j$ is preserved. Once the binary tree is created, the Dasgupta cost $C_{D}$ is calculated on the tree. A good tree with distinct clusters is characterised by a low cost $C_{D}$. Minimizing the Dasgupta cost merges similar nodes in the hierarchy, resulting in a tree with nodes clustered into appropriate communities. If $T$ is a binary tree, and $i, j$ is a pair of nodes with similarity $sim_{i, j}$ between the two nodes, HypHC minimizes the Dasgupta cost $C_{D}$ amongst all possible binary trees as shown in the equation: 
        \[T^{*} = \min_{\forall T, (i, j) \in T} C_D(T, sim_{i, j})\]
        
        With the objective function in place, HypHC minimizes the cost $C_D$ via a continuous constrained optimization problem. A continuous tree representation is built with the help of leaf nodes which are initialized with random embeddings. The leaf nodes should ultimately contain enough information to recover the full tree. All the nodes are pushed towards the boundary of a Poincar\'e disk. A Poincar\'e is the hyperbolic geometric model HypHC uses. It has a negative curvature of -1. The curvature dictates how the geometry differs from a Euclidean plane. Negative curvature makes hyperbolic manifolds behave like continuous trees. In a hyperbolic manifold, under the Poincar\'e model, the distance between two points is defined by a \textit{geodesic} \cite{hyphc_main}. 
        
        The shortest path between any two nodes must pass through their least common ancestor (LCA), which in turn aids in constructing the whole binary tree from just the boundary nodes on the Poincar\'e disk. The HypHC algorithm gives us the binary tree with minimum Dasgupta cost. From this binary tree we extract embeddings of the leaf nodes, which are our final user embeddings. Figure \ref{fig:framework} shows the application of HypHC in our work. We feed the interaction features $F$ to HypHC.
        % both lower the dimensionality and obtain community-enhanced representations of the users. 
        After reducing the dimensionality of the 600 dimensional interaction features with HypHC, we get a 60 dimensional vector for each user. 
        
    \begin{figure} [H]
        \centering
        \includegraphics[width=\linewidth]{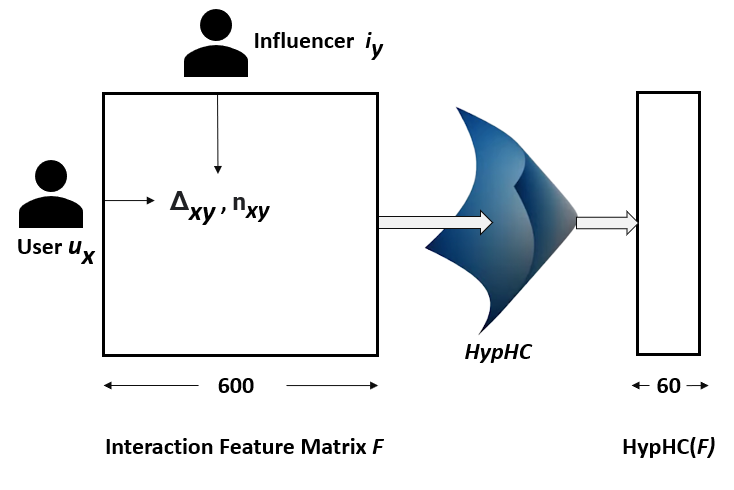}
        \caption{Overall framework: The $x$\textsuperscript{th} row in $F$ represents the interaction features $\vec{V}_x$ for a user $u_x$. The columns represent the Influencers. Each user-Influencer pair $(u_x, i_y)$ yields two features: delay $\Delta_{xy}$ and number of retweets $n_{xy}$. Thus, a total of 300 Influencers results in 600 features for each user. The dimensionality of $F$ is reduced to 60 using $HypHC()$.}
        \label{fig:framework}
\end{figure}

    % \begin{figure}
    %     \centering
    %     \includegraphics[width=150pt]{figure/hyp_hc.png}
    %     \caption{An example illustration of the intermediate continuous tree representation used in Hyperbolic Hierarchical Clustering}
    %     \label{fig:hyp_hc}
    % \end{figure}

\begingroup
\setlength{\tabcolsep}{4pt} % Default value: 6pt
% \renewcommand{\arraystretch}{1.5} % Default value: 1
% \usepackage{multirow}
\begin{table*}
\begin{subtable}{\linewidth}
\begin{tabular}{|l|cc|ccc|cc|ccc|cc|ccc|}

\hline
 & \multicolumn{5}{c|}{\textbf{Deleted vs (Suspended + Regular)}} & \multicolumn{5}{c|}{\textbf{Suspended vs (Regular + Deleted)}} & \multicolumn{5}{c|}{\textbf{Regular vs (Deleted + Suspended)}} \\ \cline{2-16} 
  % & \multicolumn{2}{c|}{\textbf{Un1ed}} & \multicolumn{3}{c|}{\textbf{Compressed}} & \multicolumn{2}{c|}{\textbf{Uncompressed}} & \multicolumn{3}{c|}{\textbf{Compressed}} & \multicolumn{2}{c|}{\textbf{Uncompressed}} & \multicolumn{3}{c|}{\textbf{Compressed}} \\ \cline{2-16} 

{\textbf{Model}} & \textbf{U} & \textbf{U+F} & \textbf{HypHC} & \textbf{SE} &
\textbf{FA} & \textbf{U} & \textbf{U+F} & \textbf{HypHC}
& \textbf{SE} & \textbf{FA} & \textbf{U} & \textbf{U+F} & \textbf{HypHC} & \textbf{SE} & \textbf{FA} \\ \hline

RFC & {\color[HTML]{212121} 90.03} & {\color[HTML]{212121} 92.25} & {\color[HTML]{212121} \textbf{93.84}} & {\color[HTML]{212121} 90.29} & {\color[HTML]{212121} 87.40} & {\color[HTML]{212121} 85.82} & {\color[HTML]{212121} \textbf{89.01}} & {\color[HTML]{212121} 87.50} & {\color[HTML]{212121} 84.23} & {\color[HTML]{212121} 84.16} & {\color[HTML]{212121} 77.31} & {\color[HTML]{212121} \textbf{79.69}} & {\color[HTML]{212121} 79.03} & {\color[HTML]{212121} 76.34} & {\color[HTML]{212121} 74.53} \\ \hline
LGBM & {\color[HTML]{212121} 90.97} & {\color[HTML]{212121} 91.14} & {\color[HTML]{212121} \textbf{92.73}} & {\color[HTML]{212121} 90.23} & {\color[HTML]{212121} 87.77} & {\color[HTML]{212121} 86.68} & {\color[HTML]{212121} \textbf{88.49}} & {\color[HTML]{212121} 87.91} & {\color[HTML]{212121} 85.77} & {\color[HTML]{212121} 83.53} & {\color[HTML]{212121} 78.18} & {\color[HTML]{212121} \textbf{79.43}} & {\color[HTML]{212121} 79.31} & {\color[HTML]{212121} 76.88} & {\color[HTML]{212121}75.62} \\ \hline
XGB & {\color[HTML]{212121} 87.17} & {\color[HTML]{212121} 88.49} & {\color[HTML]{212121} \textbf{89.16}} & {\color[HTML]{212121} 86.52} & {\color[HTML]{212121} 84.40} & {\color[HTML]{212121} 83.86} & {\color[HTML]{212121} 82.91} & {\color[HTML]{212121} 84.37} & {\color[HTML]{212121} 82.34} & {\color[HTML]{212121} \textbf{83.40}} & {\color[HTML]{212121} 76.61} & {\color[HTML]{212121} 76.87} & {\color[HTML]{212121} \textbf{78.50}} & {\color[HTML]{212121} 75.23} & {\color[HTML]{212121} 73.68} \\ \hline
GBC & {\color[HTML]{212121} 87.16} & {\color[HTML]{212121} 88.46} & {\color[HTML]{212121} \textbf{89.04}} & {\color[HTML]{212121} 86.12} & {\color[HTML]{212121} 84.99} & {\color[HTML]{212121} 83.57} & {\color[HTML]{212121} 82.90} & {\color[HTML]{212121} \textbf{83.88}} & {\color[HTML]{212121} 83.18} & {\color[HTML]{212121} 83.81} & {\color[HTML]{212121} 76.52} & {\color[HTML]{212121} 76.88} & {\color[HTML]{212121} \textbf{78.58}} & {\color[HTML]{212121} 74.98} & {\color[HTML]{212121} 74.02} \\ \hline

DNN & {\color[HTML]{212121} 72.43}
& {\color[HTML]{212121} 85.46} & {\color[HTML]{212121} \textbf{88.42}} & {\color[HTML]{212121} 87.56} & {\color[HTML]{212121} 77.72} & {\color[HTML]{212121} 81.20} & {\color[HTML]{212121} \textbf{88.03}} & {\color[HTML]{212121} 84.17} & {\color[HTML]{212121} 82.43} & {\color[HTML]{212121} 81.24} & {\color[HTML]{212121} 75.63 } & {\color[HTML]{212121} \textbf{81.57}} & {\color[HTML]{212121} 78.72} & {\color[HTML]{212121} 74.07} & {\color[HTML]{212121} 73.81} \\ \hline

LSTM & {\color[HTML]{212121}80.33} 
& {\color[HTML]{212121} 89.40} & {\color[HTML]{212121} \textbf{92.27}} & {\color[HTML]{212121} 87.13} & {\color[HTML]{212121} 76.39} & {\color[HTML]{212121}67.69} & {\color[HTML]{212121} \textbf{81.98}} & {\color[HTML]{212121} 77.81} & {\color[HTML]{212121} 76.94} & {\color[HTML]{212121} 74.97} & {\color[HTML]{212121}61.44} & {\color[HTML]{212121} 68.21} & {\color[HTML]{212121} 71.38} & {\color[HTML]{212121} \textbf{76.38}} & {\color[HTML]{212121} 74.85} \\ \hline
\end{tabular}
\caption{Results of one-vs-two class classifiers}
\label{tab:onevstwo}
 \end{subtable}\par
    \begin{subtable}{\linewidth}
\begin{tabular}{|l|cc|ccc|cc|ccc|cc|ccc|}
\hline
 & \multicolumn{5}{c|}{\textbf{Deleted vs Suspended}} & \multicolumn{5}{c|}{\textbf{Suspended vs Regular}} & \multicolumn{5}{c|}{\textbf{Regular vs Deleted}} \\ \cline{2-16} 
  % & \multicolumn{2}{c|}{\textbf{Uncompressed}} & \multicolumn{3}{c|}{\textbf{Compressed}} & \multicolumn{2}{c|}{\textbf{Uncompressed}} & \multicolumn{3}{c|}{\textbf{Compressed}} & \multicolumn{2}{c|}{\textbf{Uncompressed}} & \multicolumn{3}{c|}{\textbf{Compressed}} \\ \cline{2-16} 
{\textbf{Model}} & \textbf{U} & \textbf{U+F} & \textbf{HypHC} & \textbf{SE} & \textbf{FA} & \textbf{U} & \textbf{U+F} & \textbf{HypHC} & \textbf{SE} & \textbf{FA} & \textbf{U} & \textbf{U+F} & \textbf{HypHC} & \textbf{SE} & \textbf{FA} \\ \hline

RFC & {\color[HTML]{212121} 83.34} & {\color[HTML]{212121} \textbf{86.42}} & {\color[HTML]{212121} {85.34}} & {\color[HTML]{212121} 85.14} & {\color[HTML]{212121} 81.98} & {\color[HTML]{212121} 81.51} & {\color[HTML]{212121} \textbf{87.29}} & {\color[HTML]{212121} 86.07} & {\color[HTML]{212121} 83.33} & {\color[HTML]{212121} 82.55} & {\color[HTML]{212121} 85.50} & {\color[HTML]{212121} \textbf{88.52}} & {\color[HTML]{212121} 88.01} & {\color[HTML]{212121} 81.93} & {\color[HTML]{212121} 83.61} \\ \hline
LGBM & {\color[HTML]{212121} 83.64} & {\color[HTML]{212121} \textbf{85.80}} & {\color[HTML]{212121} {84.89}} & {\color[HTML]{212121} 84.20} & {\color[HTML]{212121} 81.87} & {\color[HTML]{212121} 85.78} & {\color[HTML]{212121} \textbf{87.27}} & {\color[HTML]{212121} 87.25} & {\color[HTML]{212121} 82.34} & {\color[HTML]{212121} 83.33} & {\color[HTML]{212121} 87.20} & {\color[HTML]{212121} 88.01} & {\color[HTML]{212121} \textbf{88.91}} & {\color[HTML]{212121} 83.25} & {\color[HTML]{212121} 84.45} \\ \hline
XGB & {\color[HTML]{212121} 81.50} & {\color[HTML]{212121} \textbf{82.35}} & {\color[HTML]{212121} {82.27}} & {\color[HTML]{212121} 80.03} & {\color[HTML]{212121} 79.11} & {\color[HTML]{212121} 82.84} & {\color[HTML]{212121} 82.37} & {\color[HTML]{212121} \textbf{84.06}} & {\color[HTML]{212121} 81.39} & {\color[HTML]{212121} 79.45} & {\color[HTML]{212121} 84.77} & {\color[HTML]{212121} 85.71} & {\color[HTML]{212121} \textbf{86.02}} & {\color[HTML]{212121} 80.89} & {\color[HTML]{212121} 81.93} \\ \hline
GBC & {\color[HTML]{212121} 81.22} & {\color[HTML]{212121} 82.33} & {\color[HTML]{212121} \textbf{82.76}} & {\color[HTML]{212121} 81.85} & {\color[HTML]{212121} 80.87} & {\color[HTML]{212121} 83.21} & {\color[HTML]{212121} 82.80} & {\color[HTML]{212121} \textbf{84.31}} & {\color[HTML]{212121} 81.43} & {\color[HTML]{212121} 79.85} & {\color[HTML]{212121} 84.56} & {\color[HTML]{212121} \textbf{84.70}} & {\color[HTML]{212121} {84.37}} & {\color[HTML]{212121} 82.15} & {\color[HTML]{212121} 81.64} \\ \hline

DNN & {\color[HTML]{212121}78.15}
& {\color[HTML]{212121} \textbf{84.46}} & {\color[HTML]{212121} 83.81} & {\color[HTML]{212121} 81.39} & {\color[HTML]{212121} 80.23} & {\color[HTML]{212121} 81.70} & {\color[HTML]{212121} \textbf{85.13}} & {\color[HTML]{212121} 83.70} & {\color[HTML]{212121} 82.42} & {\color[HTML]{212121} 81.05} & {\color[HTML]{212121} 73.36} & {\color[HTML]{212121} 81.67} & {\color[HTML]{212121} \textbf{86.59}} & {\color[HTML]{212121} 72.58} & {\color[HTML]{212121} 74.29} \\ \hline

LSTM & {\color[HTML]{212121}76.99} 
& {\color[HTML]{212121} \textbf{82.73}} & {\color[HTML]{212121} 81.17} & {\color[HTML]{212121} 78.92} & {\color[HTML]{212121} 78.52} & {\color[HTML]{212121}81.20} & {\color[HTML]{212121} \textbf{85.87}} & {\color[HTML]{212121} 83.28} & {\color[HTML]{212121} 80.23} & {\color[HTML]{212121} 81.48} & {\color[HTML]{212121} 69.76} & {\color[HTML]{212121} 80.72} & {\color[HTML]{212121} \textbf{86.38}} & {\color[HTML]{212121} 74.29} & {\color[HTML]{212121} 71.48} \\ \hline
\end{tabular}
\caption{Results of one-vs-one class classifiers}
\label{tab:onevsone}
 \end{subtable}\par
 \caption{F1 scores obtained on the classifiers and feature sets as described in Section \ref{sec:classifier}. For each of the class separation configurations, we bold the best obtained F1 score for each classifier. Observe how across all classifiers and separations, appending our proposed features to the user-level features (U+F) leads to better results than just the user-level features (U). Also observe that in almost all cases HypHC performs better than SE and FA. The F1 scores are scaled out of 100.}
 \label{result_table}
\end{table*}
\endgroup

}

\section{Results}
{
    In this section we present our experiments to evaluate the effectiveness of our features in segregating the three classes, and analysis of the same. We also evaluate the effectiveness of HypHC as a dimensionality reduction technique.
    
    \subsection{Classifier Results}    
    \label{sec:classifier}
    While comparing the different classes of users, we trained each model to classify the users in a one vs two fashion by training for \textit{suspended vs (regular + deleted)},  \textit{regular vs (deleted + suspended)} and \textit{deleted vs (regular + suspended)} separations. We also trained each model to classify the users in a one vs one (i.e. binary) fashion by training for \textit{suspended vs deleted}, \textit{suspended vs regular} and \textit{deleted vs regular} separations. 
    
    We use two types of classifier models: deep learning based and tree based. For the deep learning based classifiers, we use a deep neural network and an LSTM. The results are presented after appropriate hyperparameter tuning in the loss function, activation function, the optimizer used, number of layers and the number of epochs. For the tree based classifiers, we use lightGBM (LGBM),
    % \cite{NIPS2017_6449f44a}
    XGBoost (XGB),
    % \cite{inproceedings}
    Gradient Boosting Classifier (GBC)
    % \cite{10.3389/fnbot.2013.00021}
    and the Random Forest Classifier (RFC).
    % \cite{Statistics01randomforests}
    We use Grid Search to find the best set of hyperparameters for the tree based models. To account for class imbalance, we balance our training dataset using SMOTE \cite{Chawla_2002}.
     
    % The best test results for the various models are shown in Table \ref{result_table}.
    \subsubsection{Comparison with standard feature engineering processes: 
    }
    \label{sec:comparison_user_level}
    To evaluate the performance of our proposed features, we calculate 13 additional features for each user which are total number of tweets, number of tweets that are retweets, number of friends, number of followers, total likes, friends to follower ratio, time since account creation, lengths of screen name and bio (in characters and words) and average length of the tweet (in characters and words). We refer to these as \textit{user-level features}. These features have been used in previous works to distinguish between deleted, suspended and regular users on Twitter \cite{volkova2017identifying, runet}. We do not use psycholinguistic features for comparison for reasons discussed in Section \ref{sec:featureenginering}.
    
    To establish the benefit of our proposed features, we use various feature sets, two of which are: 
    
    \begin{enumerate}
    \item \textbf{U}: 13 dimensional user-level features as described above. 
    \item \textbf{U+F}: 613 dimensional features formed by appending the user-level features ($U$) to the 600 dimensional interaction features ($F$).
    \end{enumerate}
    
     Table \ref{result_table} clearly shows that there is an increase in F1 scores in all the cases when the user-level features are appended with the interaction features (compare columns $U$ and $U+F$), showing that our features help the various models to achieve better separation.
     
     \subsubsection{Comparison with other dimensionality reduction techniques:}
     To establish our choice of HypHC as a dimensionality reduction technique, we compared its performance with popular unsupervised dimensionality reduction methods like Principal Component Analysis (PCA), t-distributed Stochastic Neighbor Embedding (t-SNE), Spectral Embedding (SE) and Feature Agglomeration (FA). Spectral Embedding and t-SNE are dimensionality reduction techniques based on manifold learning. Feature Agglomeration applies hierarchical clustering. 
     
     We reduced the 600 dimensional features ($F$) to 30, 60, 80 and 100 dimensions using HypHC, PCA, t-SNE, SE and FA, and appended the 13 dimensional user-level features ($U$). The results observed after reducing to the dimensions mentioned above followed the same pattern -- which was that HypHC outperformed all the reduction techniques. For the sake of brevity, we only present and discuss the results obtained after reducing the dimensions to 60 (which gave the highest F1 scores overall) and with the Spectral Embedding and Feature Agglomeration techniques, which performed the best out of our comparison techniques. 
     
     We thus obtain our next set of features:
    
    \begin{enumerate}
    \item \textbf{HypHC features}: 73 dimensional features formed by reducing the 600 dimensional features ($F$) to 60 dimensions using HypHC and appending the 13 dimensional user-level features ($U$).
    \item \textbf{SE features}: 73 dimensional features formed by reducing the 600 dimensional features ($F$) to 60 dimensions using SE and appending the 13 dimensional user-level features ($U$).
    \item \textbf{FA features}: 73 dimensional features formed by reducing the 600 dimensional features ($F$) to 60 dimensions using FA and appending the 13 dimensional user-level features ($U$).
    
    \end{enumerate}

    HypHC, as mentioned in Section \ref{hypclussec}, takes advantage of hierarchical community detection to perform unsupervised dimensionality reduction to better separate classes. We evaluate the effectiveness of HypHC using two methods. First, we compare the performance of $HypHC$ features with the original $U+F$ features. We find that in most cases, $HypHC$ features perform at par with the  $U+F$ features. Moreover, there are cases where $HypHC$ features perform better than the $U+F$ features (compare columns $U+F$ and $HypHC$). This shows that HypHC is an effective dimensionality reduction technique that is able to preserve the separation between classes even at a much lower dimension.
    
    Second, we compare the performance of HypHC  with established unsupervised dimensionality reduction techniques like SE and FA. Barring a few cases in rows 3 and 6 in Table \ref{tab:onevstwo}, we find that the $HypHC$ features outperform both $SE$ features and $FA$ features. This shows that HypHC is the superior dimensionality reduction technique.
    
    \subsection{Interclass Distances}{
    \label{sec:interclass}
    To further evaluate the performance of HypHC, we compare the distances between the centers of the three classes after performing dimensionality reduction of the interaction features $F$ using HypHC, SE and FA. We take each of these three representations, and standardize each feature by removing the mean and scaling to unit variance. Table 
    \ref{distance_table} shows the cosine distances between the centers of each of the three classes in the 60 dimensional space that was generated by each of the dimensionality reduction methods. It is immediately obvious that HypHC is able to achieve the highest and most uniform separation between the classes. This ratifies our claim that HypHC is able to obtain the best separation between the classes.

    \begin{table} [H]
    \begin{tabular}{|c|r|r|r|}
    \hline
    \multicolumn{1}{|l|}{} & \multicolumn{1}{c|}{\textbf{HypHC}} & \multicolumn{1}{c|}{\textbf{SE}} & \multicolumn{1}{c|}{\textbf{FA}} \\ \hline
    \textbf{Deleted vs Suspended} & 10.9165 & 3.1636 & 2.9524 \\ \hline
    \textbf{Suspended vs Regular} & 10.9197 & 3.4279 & 2.1796 \\ \hline
    \textbf{Regular vs Deleted} & 10.9175 & 3.9857 & 3.2614 \\ \hline
    \end{tabular}
    \caption{Inter-class cosine distances obtained after reducing dimensions of the \textit{F} features to 60 dimensions with HypHC, SE and FA.}
    \label{distance_table}
    \end{table}
}

Our observations from Sections \ref{sec:classifier} and \ref{sec:interclass} show the effectiveness of our interaction features, and the valuable advantage of using HypHC in our pipeline. On comparing the results of all the classifiers for the dimensionality reduced ($HypHC$) and high dimensional ($U+F$) features, we can see that the HypHC features often outperform the original features despite being at a much lower dimension. The added bonus of using lower dimensional data is decreased storage space and lower computation cost and time. 
}

\section{Conclusion}
{

Ensuring that interactions between politicians and voters remain organic is critical to the fair functioning of any OSN, especially during democratic processes like elections. To do so, we capture these interaction patterns through our designed features. These interaction features are able to distinguish between the three classes effectively. To ensure that the model can run efficiently and take up as little space as possible, it is important to reduce the dimensionality of the features. To this end, we leverage HypHC, a novel unsupervised dimensionality reduction technique. We  show that HypHC performs better than other established dimensionality reduction techniques at separating the classes. Since our interaction features are OSN-agnostic, we plan to carry out these same analyses on other platforms.   

% Our pipeline for the extraction of interaction features provides a useful method to distinguish between different classes of users on Twitter. Our representation mechanism is made efficient by leveraging Hyperbolic Heirarchical Clustering. We successfully compress our data to 1/10th of the size while still being able to achieve F1 scores on our classifiers which are at par, and sometimes even better, with the original high dimensional data.
% In order to obtain better classifier scores, we can extract more features using the tweet content itself to append to our feature vector. Also, our features can also be used to distinguish between different classes of users in other OSNs, and to analyse other social phenomemnon on social networks. 

}
% \TK{Talk about adding more features, network and laguage features etc}
% \TK{Evaluating the performance of our features+HypHC in other settings}
% }
\vspace{-1ex}
\newcommand{\BIBdecl}{\setlength{\itemsep}{0.001 em}}
 % \bibliographystyle{IEEEtran}
\bibliographystyle{ACM-Reference-Format}
\bibliography{references}

% \renewcommand{\appendixname}{Appendix~\Alph{section}}

%update 1,2,4,8
%keep 5,6,7
%add 3
%Fig.\ref{intro} with a new element labeled “xxx” to explain xxxxx

\end{document}